\documentstyle[12pt]{article}
\begin{document}

\author{Marc Henneaux \\
{\it  Physics Department, Universit\'e
Libre de Bruxelles,} \\
{\it Campus Plaine C.P. 231,
B-1050  Bruxelles, Belgium} \\  and  \\
{\it Centro de Estudios Cient\'{\i}ficos de Santiago,} \\
{\it Casilla 16443,
Santiago 9, Chile}}

\title{On The Gauge-Fixed BRST Cohomology}
\maketitle

\begin{abstract}
A crucial property of the standard antifield-BRST cohomology at non negative
ghost number is that any cohomological class is completely determined by its
antifield independent part. In particular, a BRST cocycle that vanishes when
the antifields are set equal to zero is necessarily exact.\ \ This property,
which follows from the standard theorems of homological perturbation theory,
holds not only in the algebra of local functions, but also in the space of
local functionals. The present paper stresses how important it is that the
antifields in question be the usual antifields associated with the gauge
invariant description. By means of explicit counterexamples drawn from
the free Maxwell-Klein-Gordon system,
 we show that the property does not hold, in the case of
local functionals, if one replaces the antifields of the gauge invariant
description by new antifields adapted to the gauge fixation. In terms of
these new antifields, it is not true that a local functional weakly
annihilated by the gauge-fixed BRST generator determines a BRST cocycle; nor
that a BRST cocycle which vanishes when the antifields are set equal to zero
is necessarily exact.
\end{abstract}

\break

\section{Introduction}

Recently, the gauge-fixed BRST cohomology introduced and studied in \cite
{FH,MH2,MH1} has attracted a considerable amount of interest in the context
of the Pauli-Villars regularization of the antifield formalism \cite
{Troost,KUL1,KUL2,GomisParis1,Gomis2}.  This cohomology arises after one
has made the redefinition of the antifields appropriate to the
gauge-fixing procedure.

In  analogy with properties that are known
to hold in the standard gauge-invariant formulation of
the BRST symmetry,  one might conjecture
that any local functional $A$ annihilated by the gauge-fixed
BRST generator determines uniquely a BRST cocycle, which reduces to $A$ when
one sets the antifields associated with the gauge-fixed description equal to
zero.  If true, this conjecture would be quite useful in
many quantum calculations, which are effectively performed after the gauge
is fixed.

The purpose of this paper is to clarify the structure of the gauge-fixed
BRST cohomology in the space of local functionals. We show by means of
explicit counterexamples that the above conjecture
unfortunately does not hold.\ \ We also explain why this is not so. The
difficulties arise because the differential that occurs in
the gauge-fixed perturbation expansion (Koszul differential
associated with the gauge-fixed equations of motion) is not acyclic
in the space of local functionals, {\it even at positive
antifield number and positive ghost
number}.  This is in sharp contrast with what happens in the
usual gauge-invariant formulation, and results from the fact
that the ghosts are subject to non trivial equations of motion
once the gauge is fixed.  Consequently, one can and does meet
obstructions in trying to reconstruct the BRST-invariant extension of a given
cocycle of the gauge-fixed cohomology.  And furthermore, there
exists non trivial cocycles that vanish when one sets
the redefined antifields equal to zero.
This implies  that a BRST cohomological class in the space of local
functionals is not uniquely determined by its antifield independent
part in the gauge-fixed expansion.
 From the mathematical point of view, the gauge fixed
perturbation expansion in the redefined antifields
provides an example of homological perturbation theory with
a non acyclic differential \cite{HT}.

\section{Standard BRST Cohomology}

We first briefly recall some of the salient points of the antifield
formalism necessary for the subsequent discussion.

The antifield formulation of the BRST theory \cite{BRS,T} appears to be one
of the most powerful tools for quantizing systems with a gauge freedom \cite
{ZJ,Ka,dWvH,BV,VT}.\ \ In that approach, one replaces the original gauge
invariant action $S_0[\Phi ^i]$ depending on the classical fields $\Phi ^i$
by the so-called minimal solution of the master equation,
\begin{equation}
S=S_0+\int dx\int dy\ \Phi _i^{*}(x)R_{\ \alpha }^i(x,y)C^\alpha (y)+\hbox{
``more''}  \label{classicalS}
\end{equation}
where :

\noindent (i) $\Phi _i^{*}$ are the antifields associated with the fields $%
\Phi ^i$;

\noindent (ii) $C^\alpha $ are the ghosts;

\noindent (iii) $\delta _\varepsilon \Phi ^i(x)=\int dy\ R_{\ \alpha
}^i(x,y)\varepsilon ^\alpha (y)$ are the gauge symmetries, under which the
action is invariant, $\delta _\varepsilon S_0=0$; the bilocal objet $R_{\
\alpha }^i(x,y)$ is a finite sum of terms, each of which involves $\delta
(x,y)$ or one of its derivatives, $R_{\ \alpha }^i(x,y)=\sum B_{(m)}\partial
^{(m)}\delta (x,y)$; the coefficients $B_{(m)}$ depend on the fields at $x$,
say, and their derivatives up to a finite order;

\noindent (iv) ``more'' in (\ref{classicalS}) depends on the fields $\Phi
^A\equiv (\Phi ^i,C^\alpha )$ and their conjugate antifields $\Phi
_A^{*}\equiv (\Phi _i^{*},C_\alpha ^{*})$, contains terms
proportional to $(\Phi^*_i)^p (\Phi^*_\alpha)^q$ with $p+2q>1$,
and is adjusted so that $S$ solves the classical master equation,
\begin{equation}
(S,S)=0.  \label{master}
\end{equation}

We assume for simplicity that the gauge symmetry is irreducible but our
consideration apply equally well to reducible gauge symmetries. We also
assume that the reader is familiar with the basic principles of the
antifield formalism. Besides the original work quoted above, the reader may
consult the references \cite{MH1,HT,Gomis} for further information. We shall
follow in particular the approach of \cite{MH1,HT}, based itself on \cite
{FH,HT2,FHST}, where the rationale of the antifield formalism is explained
from the cohomological point of view. This approach turns out to be crucial
for understanding the difficulties associated with the gauge-fixed
cohomology.

Because $S$ solves the classical master equation, the left derivation
defined by the equation
\begin{equation}
sA=(S,A)  \label{BRST}
\end{equation}
is a differential,
\begin{equation}
s^2=0,
\end{equation}
the so-called BRST differential. One can thus define the BRST cohomological
groups $H(s)$ in the standard manner, as the quotient spaces $Ker(s)/Im(s)$
of the BRST cocycles modulo the BRST coboundaries. Explicitly, the cocycle
condition reads
\begin{equation}
sA=0\hbox{ (}s\hbox{-cocycle condition),}  \label{scocyc}
\end{equation}
while the coboundary condition is
\begin{equation}
A=sB\hbox{ (}s\hbox{-coboundary condition).}  \label{scobound}
\end{equation}
One can actually consider various BRST cohomologies, depending on which
functional space $A$ and $B$ are required to belong to. The following
choices are met in practice :

\noindent (i) formal BRST cohomology $H^{\hbox{formal}}(s)$, in which $A$
and $B$ are allowed to be arbitrary functionals;

\noindent (ii)BRST cohomology $H^{\hbox{loc.funct.}}(s)$ in the space of
local functionals, in which $A$ and $B$ are required to be local
functionals, i.e., integrals of local functions (a local function is a
function of the fields $\Phi ^A$, the antifields $\Phi _A^{*}$ and a finite
number of their derivatives);

\noindent (iii)BRST cohomology $H^{\hbox{local}}(s)$ in the space of local
functions, in which $A$ and $B$ are required to be local functions.

To analyse the BRST cohomology, it is convenient to introduce the antighost
number, the pure ghost number and the (total) ghost number as follows,
\begin{eqnarray}
\hbox{antigh}(\Phi ^i) &=&0\hbox{ , pure gh}(\Phi ^i)=0\hbox{ , gh}(\Phi
^i)=0\hbox{ ,} \\
\hbox{antigh}(C^\alpha ) &=&0\hbox{ , pure gh}(C^\alpha )=1\hbox{ , gh}%
(C^\alpha )=1\hbox{ ,} \\
\hbox{antigh}(\Phi _i^{*}) &=&1\hbox{ , pure gh}(\Phi ^i)=0\hbox{ , gh}(\Phi
_i^{*})=-1\hbox{ ,} \\
\hbox{antigh}(C_\alpha ^{*}) &=&2\hbox{ , pure gh}(C_\alpha ^{*})=0\hbox{ ,
gh}(C_\alpha ^{*})=-2\hbox{ .}
\end{eqnarray}
The ghost number is the difference between the pure ghost number and the
antighost number.

Given $S$, one can expand the differential $s$ according to the antighost
number. One gets
\begin{equation}
s=\delta +\gamma +\sum_{k\geq 1}s_k\hbox{, antigh}(\delta )=-1\hbox{, antigh}%
(\gamma )=0\hbox{, antigh}(s_k)=k\hbox{.}  \label{HPT1}
\end{equation}
The nilpotency of $s$ implies the following relations,
\begin{equation}
\delta ^2=0,\quad \delta \gamma +\gamma \delta =0,\quad \gamma ^2+(\delta
s_1+s_1\delta )=0,\hbox{ etc.}  \label{HPTexp}
\end{equation}
The first term in the expansion of $s$ is the Koszul-Tate differential
associated with the gauge-invariant equations of motion.
It plays a central role in BRST theory. The second term
is the longitudinal exterior derivative along the gauge orbits and
is related to the gauge symmetry \cite{FH,MH1,HT}.

It is clear that if $A$ is a BRST
cocycle and has non negative antighost number, then its component $A_0$
independent of the antifields fulfills the condition
\begin{equation}
sA=0\Rightarrow \gamma A_0+\delta A_1=0,  \label{HPT2}
\end{equation}
where
\begin{equation}
A=A_0+\sum_{k\geq 1}A_k\hbox{, antigh}(A_k)=k.
\end{equation}
Conversely one has

\vspace{.5cm}

\noindent
{\bf Theorem 1 : } Any solution $A_0$
of (\ref{HPT2}) determines a unique BRST cohomological
class.\ \

\vspace{.5cm}

\noindent
{\bf Theorem 2 : } Any BRST cocycle $A$
with non negative ghost number that vanishes when the
antifields are set equal to zero ($A_0=0$) is BRST-exact, $A=sB$.

\vspace{.5cm}

These theorems hold equally well for $H^{\hbox{formal}}(s)$, $H^{\hbox{%
loc.funct.}}(s)$ and $H^{\hbox{local}}(s)$ and are direct consequences of
the perturbation expansion (\ref{HPT1}) and of the acyclic properties of the
Koszul-Tate differential. They are proved in \cite{HT2,FHST,DV} in the
Hamiltonian framework and in \cite{FH,MH1} in the antifield case (see also
\cite{HT}, chapter 8, section 8.4 - in particular page 181 - for a general
perspective independent of the precise context).

Since  $\delta A_1$ in (\ref{HPT2}) is proportional to the equations
of motion, the antifield independent component $A_0$ of
$A$ is a $\gamma$-cocycle on-shell, i.e., is in the
weak cohomology of the longitudinal derivative $\gamma$.  Thus the
theorems
state in fact that any cohomological class of the weak longitudinal
cohomology $H^{\hbox{weak}}(\gamma)$ determines uniquely a BRST cohomological
class.  Any representative of this BRST cohomological class is
called a ``BRST-invariant extension'' of $A_0$ \cite{HT2,FHST,FH,MH1,HT}.

We shall not repeat the proof here, but shall only sketch the central idea
by recalling how one reconstructs $A$ from $A_0$. Since $H^{\hbox{loc.funct.}%
}(s)$ is the trickier case, we shall from now on restrict the analysis to
local functionals, $A=\int a$, where $a$ is a local $n$-form. In terms of the
integrands, the $s$-cocycle and $s$-coboundary conditions (\ref{scocyc}) and
(\ref{scobound}) become respectively
\begin{eqnarray}
sa &=&dm\hbox{~(}s\hbox{-cocycle condition),}  \label{scocyc'} \\
a &=&sb+dn\hbox{ (}s\hbox{-coboundary condition),}  \label{scobound'}
\end{eqnarray}
for some $m$ and $n$ since the integral of a $d$-exact $n$-form is a surface
term and vanishes (we assume appropriate boundary conditions). For
this reason, one can identify the cohomological group
$H^{\hbox{loc.funct.}}(s)$
with the cohomological group $H(s|d)$ of $s$ modulo $d$ in the space
of local $n$-forms.

In the same way,
the equation (\ref{HPT2}) on $A_0$ becomes
\begin{equation}
\gamma a_0+\delta a_1=dm_0.  \label{HPT2'}
\end{equation}
To reconstruct $a$ from a solution $a_0$ of (\ref{HPT2'})
(for some $a_1$ and $%
m_0$), one proceeds as follows : first, one notes that (\ref{HPTexp}) and (%
\ref{HPT2'}) imply that $b_1\equiv \gamma a_1+s_1a_0$ is a $\delta $-cycle
modulo $d$, $\delta b_1=dp$ for $p=\gamma m_0$. Furthermore, $b_1$ has
antighost number $1$.
In order to be able to complete $a$ into an $s$%
-cocycle modulo $d$, it is necessary that $b_1$ be a $\delta $-boundary
modulo $%
d$, $b_1+\delta a_2=dm_1$ for some $a_2$ and $m_1$. This ensures that $%
a_0+a_1+a_2$ fulfills the condition $sa=0$ (modulo $d$) not just at
antighost number zero - as implied by (\ref{HPT2'}) - but also at antighost
number one. One can then adjust successively $a_3$, $a_4$ etc in such a way
that $a=\sum_ka_k$ is a $s$-cocycle modulo $d$ to all orders.  One says
that the construction of $a$ is not obstructed.

The question boils down therefore to the question of whether the
cohomological groups $H_k(\delta |d)$ in which the
potential obstructions could lie vanish for $k>0$, in the space of
local $n$-forms : is any $\delta $-cycle modulo $d$ with strictly positive
antighost number automatically $\delta $-exact modulo $d$? As shown in \cite
{MH3} by means of an explicit counterexample, the answer to this question is
negative.\ \ There exist obstructions, which have been related in \cite{BBH1}
to the characteristic cohomology \cite{BryantG} (conserved currents,
conserved $p$-forms) of the gauge-invariant field equations. However, these
obstructions are not met in the above reconstruction process because $b_1$
(and the successive $b_k$'s) has strictly positive pure ghost number. As
proved  in \cite{MH3}, the cohomological groups $H_k(\delta |d)$ vanish for $%
k>0$ {\bf and} strictly positive pure ghost number. The reason for this is
that the ghosts are subject to no equations of motion in the gauge invariant
formulation of the theory.\ \ More precisely, they are free in the homology
of $\delta $ : there is no non-trivial relation among the ghosts that can be
written as a $\delta $-boundary. This property is crucial and enables one to
avoid the obstructions that one could otherwise meet in the reconstruction
of $a$ from $a_0$.

As we shall now discuss, there is no such equivalent
property in the gauge fixed formulation of the theory.  Accordingly,
there is no
analog of Theorems 1 and 2 above for the
gauge-fixed cohomology in the space of
local functionals. One may (and does) meet obstructions because the ghosts
are subject to equations of motion.  The differential arising in
the perturbative reconstruction of the BRST cocycles in
the gauge-fixed description is not acyclic, even
when restricted to the degrees that occur in the expansion.

\section{Gauge-Fixed BRST\ Cohomology}

In order to quantize the theory, it is necessary in practice to fix the
gauge.\ \ To that end, one introduces non-minimal variables. The most common
choice is to add antighosts $\overline{C}_\alpha $ and Nakanishi-Lautrup
auxiliary fields $b_\alpha $, with BRST transformation laws
\begin{equation}
s\overline{C}_\alpha =b_\alpha \hbox{ , }sb_\alpha =0\hbox{.}
\label{nonmins}
\end{equation}
The corresponding antifields are denoted by $\overline{C}^{*\alpha }$ and $%
b^{*\alpha }$. The transformation (\ref{nonmins}) is obtained by adding the
term $-\int dx\ b_\alpha \overline{C}^{*\alpha }$ to $S$,
\begin{equation}
S\rightarrow S-\int dx\ b_\alpha \overline{C}^{*\alpha }\hbox{ ,}
\end{equation}
which preserves the master equation.\ \ Unless otherwise specified, the
fields $\Phi ^A$ will now collectively refer to the original classical
fields, the ghosts, the antighosts and the auxiliary $b$-fields, while the
antifields $\Phi _A^{*}$ will stand for all the conjugate antifields.

One
then eliminates the antifields in a two-step procedure :

\noindent (i) First, one makes the canonical transformation
\begin{equation}
\Phi ^{\prime A}=\Phi ^A\hbox{ , }\Phi _A^{\prime *}=\Phi _A^{*}-\frac{%
\delta \Psi }{\delta \Phi ^A}\hbox{ }
\end{equation}
where the odd functional $\Psi $ is called the gauge-fixing fermion;

\noindent (ii) Second, one sets the new antifields $\Phi _A^{\prime *}$
equal to zero. If $\Psi $ is well chosen, the resulting gauge-fixed action $%
S_\Psi $,
\begin{equation}
S_\Psi [\Phi ^A]=S[\Phi ^A,\Phi _A^{*}=\frac{\delta \Psi }{\delta \Phi ^A}]%
\hbox{ ,}  \label{gfxaction}
\end{equation}
leads to non-degenerate equations of motion, i.e., has no residual gauge
invariance.

In discussing the gauge-fixed formulation of the theory, it is convenient to
introduce a new degree, called the antifield number $r$, which puts all the
antifields on an equal footing \cite{MH2},
\begin{equation}
r(\Phi ^A)=0,\quad r(\Phi _A^{\prime *})=1.
\end{equation}
One can expand the BRST differential and the BRST cocycles according to this
new degree,
\begin{eqnarray}
s &=&\delta _\Psi +\gamma _\Psi +\sum_{k\geq 1}s_k^{\prime },~r(\delta _\Psi
)=-1,~r(\gamma _\Psi )=0,~r(s_k^{\prime })=k,  \label{HPTanf} \\
A &=&\sum_kA_k^{\prime },\quad r(A_k^{\prime })=k.  \label{HPTanf2}
\end{eqnarray}
The differential $\delta _\Psi $ is the Koszul differential associated with
the gauge-fixed equations of motion \cite{MH2}.\ \ It is acyclic in positive
antifield degree in the space of formal functionals as well as in the
algebra of local functions. However, it is not acyclic in the space of local
functionals, the obstructions being related to the global symmetries
(conserved currents) of the gauge-fixed action. Furthermore, one cannot use
the pure ghost number as an auxiliary tool for controlling the cohomology,
because the ghosts are now subject to non trivial equations of motion. For
this reason, Theorems 1 and 2 of the previous section apply to the expansion
according to the antifield number (\ref{HPTanf}) and (\ref{HPTanf2}) if one
deals with the formal BRST cohomology and the local BRST cohomology, but
{\bf not} in the case of the BRST cohomology in the space of local
functionals.

We shall exhibit explicit counterexamples in the next section. Before doing
this, we note that the action of $\gamma _\Psi $ on the fields can be
written as
\begin{equation}
\gamma _\Psi \Phi ^A=(S,\Phi ^A)|_{\Phi _A^{*}=\frac{\delta \Psi }{\delta
\Phi ^A}}\hbox{ .}
\end{equation}
One easily verifies that $(\gamma _\Psi )^2$ is weakly zero on the fields,
\begin{equation}
(\gamma _\Psi )^2\approx ^{\prime }0
\end{equation}
where $\approx ^{\prime }$ means ``equal modulo the equations of motion $%
\delta S_\Psi /\delta \Phi ^A=0$ following from the {\em gauge-fixed}
action'', and that the gauge-fixed action (\ref{gfxaction}) is invariant
under the gauge-fixed BRST symmetry $\gamma _\Psi $. One can then define the
weak BRST cohomology $H^{\hbox{weak}}(\gamma _\Psi )$ in the space of
function(al)s of the fields by the conditions,
\begin{eqnarray}
\gamma _\Psi A[\Phi ^A] &\approx ^{\prime }&0\hbox{ (}\gamma _\Psi \hbox{%
-cocycle condition)}  \label{gfxcocyc} \\
A[\Phi ^A] &\approx ^{\prime }&\gamma _\Psi B[\Phi ^A]\hbox{ (}\gamma _\Psi
\hbox{-coboundary condition).}  \label{gfxcobound}
\end{eqnarray}
Again, one may consider various cases, depending on the functional spaces to
which $A$ and $B$ belong . Since $sA=0$ implies $\gamma _\Psi A_0^{\prime
}\approx ^{\prime }0$ for the component of $r$-degree zero of $A$, one
sometimes refers to Theorems 1 and 2 in the context of the gauge-fixed
expansion as the reconstruction theorems for the gauge-fixed cohomology \cite
{Gomis2}. Our main result is thus that these reconstruction theorems do not
hold for local functionals.

\section{The Counterexamples}

\subsection{The Model}

The counterexamples discussed here
arise in the case of the combined free
Maxwell-Klein-Gordon system, with a massless scalar field. The
gauge-invariant action is quadratic and equal to the sum of the free Maxwell
action and the free KG action for a neutral (real) scalar field,
\begin{equation}
S_0[A_\mu ,\phi ]=\int dx\ \left[ -\frac 14F_{\mu \nu }F^{\mu \nu }-
\frac12\partial
^\mu \phi \partial _\mu \phi \right]
\end{equation}
In the Lorentz gauge enforced through the standard Gaussian average choice
of gauge-fixing fermion, the gauge-fixed action is, with appropriate sign
and factor conventions,
\begin{equation}
S=\int dx\left[ -\frac 14F_{\mu \nu }F^{\mu \nu }+
\partial _\mu \overline{C}\partial ^\mu C +
b (\partial_\mu A^\mu + \frac12 b) -
\frac12 \partial
^\mu \phi \partial _\mu \phi\right]
\end{equation}
The BRST symmetry reads $s = \delta_\Psi + \gamma_\Psi$, with
\begin{equation}
\gamma_\Psi A_\mu = \partial_\mu C, \; \gamma_\Psi \phi = 0,\;
\gamma_\Psi C = 0, \; \gamma_\Psi b =0, \; \gamma_\Psi \overline{C} = b
\end{equation}
and
\begin{equation}
\gamma_\Psi A'^{* \mu} = 0, \; \gamma_\Psi \phi'^* = 0, \;
\gamma_\Psi C'^* = - \partial_\mu A'^{*\mu},
\; \gamma_\Psi b'^* = - \overline{C}'^*, \; \gamma_\Psi
\overline{C}'^* = 0.
\end{equation}
The Koszul differential $\delta_\Psi$ associated with the gauge-fixed
stationary surface is given by
\begin{equation}
\delta_\Psi \Phi^A = 0, \; \delta_\Psi A'^{*\mu} = \partial_\nu
F^{\nu \mu} - \partial^\mu b, \; \delta_\Psi \phi'^* = \partial_\mu
\partial^\mu \phi
\end{equation}
and
\begin{equation}
\delta_\Psi C'^* = -\partial_\mu \partial^\mu
\overline{C}, \; \delta_\Psi b'^* = \partial_\mu
A^\mu + b, \; \delta_\Psi\overline{C}'^* =  \partial_\mu \partial^\mu C.
\end{equation}

\subsection{Counterexample To Reconstruction Theorem}

The first counterexample is a counterexample to the reconstruction
theorem.  Consider
the ghost number zero local function
\begin{equation}
-\frac12 A^\mu A_\mu + \overline{C} C
\end{equation}
It is a $\gamma_\Psi$-cocycle modulo $d$ on the surface of the gauge-fixed
equations of motion since
\begin{equation}
\gamma_\Psi (-\frac12 A^\mu A_\mu + \overline{C} C) +
\delta_\Psi (-b'^* C) = \partial_\mu (- C A^\mu).
\end{equation}
Furthermore, it is non trivial, i.e. it is not
weakly $\gamma_\Psi$-exact modulo $d$.  If the
reconstruction theorem were  correct, one could construct
an $s$-cocycle modulo $d$ that starts like $a = a_0 + a_1 +
a_2 + \dots$, with $a_0 = -\frac12 A^\mu A_\mu + \overline{C} C$
and $a_1 = -b'^* C$.

However, it is easy to see that there
is no $a_2$ such that $sa = 0$ (modulo $d$) up to terms of
antifield degree two.  Indeed, one has
\begin{equation}
\gamma_\Psi a_1 = \gamma_\Psi (-b'^* C)
= \overline{C}'^* C.
\end{equation}
Although $\overline{C}'^* C$ is $\delta_\Psi$-closed modulo $d$,
it is not $\delta_\Psi$-exact modulo $d$, because any
$\delta_\Psi$-boundary modulo $d$ necessarily vanishes
when one sets all the derivatives of the fields and
the $b$-field equal to zero.  Thus, the
construction of $a_2$ is explicitly obstructed, in this case
by the generator of the global symmetry $\overline{C} \rightarrow
\overline{C} + \eta C$ ($\eta$ anticommuting constant). There
is no $a$ which is BRST-closed modulo $d$ and which starts like
$a_0$, even though $a_0$ is a cocycle of the gauge-fixed BRST
cohomology. [One may easily check that the ambiguities in
$a_0$ and $a_1$ cannot be used to remove the obstruction].
Note that $\overline{C}'^* C$ has antifield number and ghost
number both equal to one.

\subsection{Counterexample To Theorem 2}

The second example is a counterexample to Theorem 2.
Let $a$ be given by
\begin{equation}
a = \overline{C}'^* \phi - \phi'^* C
\end{equation}
This object is a BRST cocycle modulo $d$, has ghost number zero and
is linear, homogeneous
in the redefined antifields associated with the gauge-fixed
description.  Thus, it vanishes if one sets
the redefined antifields $\overline{C}'^*$ and $\phi'^*$ equal to
zero.  Yet, $a$ is {\bf not} a BRST coboundary.  The
easiest way to see this is to rewrite $a$ in terms of the
original antifields.  One has
$\overline{C}'^* = \overline{C}^* + \partial_\mu A^\mu
+ (1/2) b$, $\phi'^* = \phi^*$
and thus $a = A_\mu j^\mu - \phi^* C +
s(-b^* \phi + (1/2) \overline{C} \phi + \partial_\mu (A^\mu \phi))$,
where $j^\mu$ is the non trivial conserved current $-\partial^\mu
\phi$.
Modulo an exact term, the cocycle $a$ is a non trivial
antifield-dependent
cocycle of the
form investigated in \cite{BHprl,BBH2}.  This shows that there is
non trivial BRST cohomology in the space of local functionals at non
vanishing antifield number.

Another counterexample to Theorem 2, this time at ghost number one,
can be constructed along the same lines by taking as gauge-fixing
fermion $\Psi = \overline{C} (\partial_\mu A^\mu +(1/2) b + \phi)$,
which is permissible.  The $s$-cocycle modulo $d$ given by
$\overline{C}'^* C$ reads, in terms of the original antifields,
$\overline{C}'^* C = \phi C + $  trivial terms, since
$\overline{C}'^*$ is now given by  $\overline{C}'^*
= \overline{C}^* + \partial_\mu A^\mu +(1/2) b +
 \phi$.  It is clear that $\phi C$ is a non-trivial $s$-cocycle
modulo $d$, even though $\overline{C}'^* C$
is linear, homogeneous in the redefined antifields and of
ghost number one.  Note that this counterexample does not assume the
existence of conserved currents.

\section{Conclusions}

We have proved  in this letter by means of explicitit counterexamples
that important properties of the BRST construction
that hold in the standard gauge invariant description
no longer hold when one reexpresses the theory in terms of the
redefined antifields associated with the gauge fixation.  This
is because the redefined Koszul differential is no longer
acyclic in the space of local functionals - even though it
remains acyclic, of course, in the formal space of all
functionals and in the algebra of local functions.  The
gauge-fixed description provides an example of homological
perturbation theory with a non acyclic differential, in
which one can - and does - meet obstructions.
It follows in particular from our analysis that one cannot,
in general,
determine a BRST cohomological class (e.g. anomalies at
ghost number one) by computing only its antifield independent
part in terms of the {\it redefined} antifields
although
this can be done in terms of the standard antifields.

\section{Acknowledgements}

The author is grateful to Glenn Barnich and
Joaquim Gomis for useful conversations.
This work has been supported in part by research
contracts with the F.N.R.S. and  with the Commission of the European
Community.

\end{document}